\documentstyle[prl,twocolumn,aps]{revtex}

\begin{document}

\title{ Atomic resolution STM imaging of a twisted
single-wall carbon nanotube}

\author{W. Clauss\footnote[1]{ permanent address: Institute
for Applied Physics, University of Tuebingen, Germany,
e-mail clauss@tuebingen.de}, D.J. Bergeron, and A.T.
Johnson } 

\address{Department of Physics and Astronomy, University
of  Pennsylvania, 209 S. 33rd St., Philadelphia, PA  19104,
USA}

\date{\today}
\maketitle

\begin{abstract}
We present atomically-resolved STM images of single-wall
carbon nanotubes (SWNTs) embedded in a crystalline nanotube
rope. Although they may be interpreted as of a chiral
nanotube, the images are more consistently explained as an
achiral armchair tube with a quenched twist distortion. The
existence of quenched twists in SWNTs in ropes might explain
the fact that both as-grown bulk nanotube material and
individual ropes have insulator-like conductivity at low
temperature. 

\end{abstract}


\pacs{61.48+c, 61.16Ch, 68.35.Bs}

Nanostructured materials are an intriguing laboratory for
investigating the properties of electrons and phonons in
systems intermediate between solids and single atoms. The
novel properties of \(sp^{2}\)-bonded elemental carbon make
single-walled carbon nanotubes (SWNTs)
\cite{Iijima,Ajayan97} a particularly promising
nanostructure with a remarkable number of proposed
electronic, chemical, sensing, and mechanical applications.
Since macroscopic amounts of atomically flawless SWNTs form
spontaneously in a plasma containing appropriate catalyst
atoms \cite{Thess96}, it is likely that some of these uses
will soon be realized. The bulk SWNT material produced by
present methods is an entangled mat of SWNT bundles (known
as ropes) made of tubes packed into polycrystalline
triangular arrays. 

Well before they could be produced in quantity, it was
appreciated that the electronic nature of SWNTs is
controlled by the precise wrapping of the graphite sheet,
due to the latter's critical band structure, poised
delicately between metal and semiconductor
\cite{Hamada92,Saito92}. Later work demonstrated that tube
curvature and shape fluctuations have a strong impact on
the electronic and transport properties of SWNTs
\cite{Kane97,KMF}. Even small twists strongly backscatter
electrons propagating along a tube, while bends in the tube
act as much weaker scatterers. 

Resistivity measurements of bulk SWNT material and
individual ropes agree in large degree with calculations of
the electron scattering due to thermally excited twists of
intrinsically metallic tubes \cite{KMF}. At elevated
temperature, both bulk and rope resistivities increase
linearly with temperature as predicted, but there is a
still-mysterious transition to insulator-like behavior (i.\
e.\, \(dR/dT < 0 \)) below a crossover temperature that
depends on sample preparation. It has been suggested
\cite{Kane97} that the source of this crossover could be
the onset of backscattering from twist quenched in the
samples.

Structural and electronic investigations of SWNTs provide
little insight on this issue so far. Extensive TEM
\cite{Nikolaev97}, X-ray diffraction \cite{Thess96},
electron nano-diffraction \cite{Cowley97} and Raman
scattering \cite{Rao97} measurements, among other methods,
suggest that the samples used in our work, produced by
laser ablation at $1200^{o}$C \cite{Guo95}, contain a
significant fraction of achiral armchair tubes with
diameter near 1.4 nm. Because of their limited spatial
resolution, these techniques give little information about
the precise atomic configuration of individual tubes.
Scanning tunneling microscopy can address this issue 
\cite{Sattler95,DekLieb}, but to date no experiments have
shed light on the atomic configuration of SWNTs packed into
ropes, where tube-tube interactions likely play an important
role. 

In this paper, we present images of SWNTs within a rope
with sufficient spatial resolution to determine the
wrapping vector \cite{Saito92} and for the first time
structural distortions that we attribute to forces exerted
by their neighbors in the lattice. Our STM images could be
ascribed to a chiral nanotube, but are more consistently
explained as an armchair tube with a quenched twist
distortion. We argue that the amount of observed twist can
be qualitatively explained within existing theories of SWNT
mechanics. Our data is evidence that quenched twists exist
in SWNTs in ropes and may contribute to the observed
insulator-like conductivity at low temperature. 

In order to get well-defined, atomically flat substrates
for high-resolution STM images, we evaporate gold
onto mica substrates at $300^{o}$C. After annealing in a
gas flame, atomically flat Au(111) terraces several hundred
nanometers in size are obtained. As-grown nanotube
material was sonicated for 30 minutes in dichloroethane,
then a small drop of the solution was put on the gold
surface and blown dry in a nitrogen stream.

Measurements were taken in air with commercial Pt-Ir tips
and an STM of the Besocke design (the Beetle STM from
Omicron, Inc.). Although the instrument has a rather small
scan size (about 2 $\mu$m, it allows a reliable coarse
lateral movement with steps in the range of $50$ to $200$
nm, making it possible to find useful sample regions in a
few minutes, even with low SWNT density. Lateral dimensions
were calibrated to an estimated accuracy better than 15 \%
using atomically-resolved images of Au(111),
highly-oriented pyrolytic graphite (HOPG), and decanethiol
monolayers on Au(111). The vertical scale was calibrated at
monoatomic steps between terraces of Au(111). Images were
obtained with tunneling voltages in the range of $700-900$
mV and setpoint currents between $350$ and $900$ pA. No
significant influence of the voltage polarity was noticed. 

Figure \ref{fig1} shows a densely packed rope with a
rectangular cross-section. It is unclear whether this shape
is due to the rope-substrate interaction or if rectangular
ropes with such large aspect ratios form during the growth
process \cite{jacques}. At the side-wall of the rope,
distinct steps and terraces are seen.  A careful
examination reveals that the step heights are rather
well-quantized, implying that the rope consists of tubes
with a narrow diameter distribution. This image also lets
us accurately determine the direction of the tube axes.

All high resolution images discussed below were taken with
the fast scan direction nearly parallel to the tube axis.
This minimizes the effect of lateral tip-sample forces that
are found to disrupt the binding between tubes and
substrate. Moreover, the amplitude of height changes during
each line scan is minimized, resulting in reduced feedback
error. 

Figure \ref{fig2}a shows a high resolution image of an area
on top of the rope. This image combines data from two
sequential scans, a procedure made possible by the STM
system's low drift. A cylindrical tube is visible, embedded
in a flat, disordered surface that may be conducting
amorphous carbon. On the surface of the tube we resolve
individual hexagons of carbon atoms arranged in the
honeycomb lattice of a single sheet of graphite. This
honeycomb pattern differs distinctly from the hexagonal
lattice typical of STM images of HOPG, where the tip senses
only every second atom in a unit cell (the one lying above
an atom in the second layer of the AB stacked crystal). The
distance between the centers of the carbon rings is $0.25$
nm, as expected for graphite, confirming the claim that we
image each atom on the tube. 

Figure \ref{fig2}b shows raw topographic data along a line
perpendicular to the tube axis. The deviation from a
circular cross-section likely results from a slightly
asymmetric tip shape. The visible tube width is
$1.4$ nm, and its height above the amorphous background is
$0.35$ nm. Assuming a circular tube cross-section,
this gives a measured tube diameter of $1.7$ nm. X-ray
diffraction data \cite{Thess96} show a rope lattice spacing
of $1.7$ nm and a peak in the SWNT diameter distribution at
$1.4$ nm.  Since STM probes the electronic rather than
nuclear tube diameter, it is not surprising that our
measured diameter is close to the rope lattice spacing,
since both include a contribution from the electron cloud
surrounding the tube. 

Following Refs.\ \onlinecite{Hamada92,Saito92}, we index a
SWNT by a pair of integers $(n,m)$ corresponding to the
wrapping vector \( \mbox{\boldmath$L$} = n
\mbox{\boldmath$a_{1}$} + m \mbox{\boldmath$a_{2}$} \) that
defines the elementary orbit around the tube waist. We also
define the armchair and zig-zag directions $(n,n)$ and
$(n,0)$. Each nanotube index $(n,m)$ has a characteristic
diameter and chiral angle between the zig-zag direction and
the tube axis. For example, the chiral angle of an
undistorted armchair tube (see Fig.\ \ref{fig3}a) with
wrapping vector $(n,n)$ is zero, since the armchair (solid
lines) and zig-zag (horizontal dotted arrow) directions are
perpendicular and parallel to the tube axis, respectively.
Two sets of dotted parallel lines in Fig.\ \ref{fig3}a show
the other zig-zag directions.

We can take the data from Fig.\ \ref{fig2}a and suppress
the tube curvature for enhanced atomic contrast. This
yields Fig.\ \ref{fig4}, where we indicate the tube axis
and armchair and zig-zag directions as in Fig.\
\ref{fig3}a. In Fig.\ \ref{fig4} there is a chiral angle of
about $4^{o}$ between the axis and the zig-zag direction,
which, combined with the $1.4$ nm diameter found above,
gives a tentative identification of a $(12,9)$ tube
(diameter $1.43$ nm, chiral angle $4.7^{o}$). Although this
conclusion may be correct, it does not account for all
features of the image. Closer inspection of Fig.\
\ref{fig4} reveals that the armchair direction is on
average perpendicular to the tube axis, as expected for a
\emph{non-chiral} armchair tube. In addition, the average
angle between the armchair and zig-zag directions is
greater than $90 ^{o}$, implying that the tube is
\emph{distorted} from its equilibrium conformation.

All these observations can be explained if the images are
in fact of an armchair tube with a twist
distortion of $4^{o}$. The observed diameter near $1.4$ nm
would make it a $(10,10)$ tube, although the $(11,11)$
diameter is within experimental error. As sketched in Fig.\
\ref{fig3}b, a twist deformation of an armchair nanotube
causes the angle between the zig-zag and armchair
directions to differ from $90^{o}$, while the armchair
direction remains perpendicular to the tube axis, just as
in Fig.\ \ref{fig4}. Such distortions are particularly
important to the electronic properties of the tube (and
rope). A localized twist is a strong source of electron
backscattering, and a uniformly twisted SWNT has a gap at
the Fermi energy, even if the undistorted tube is metallic
\cite{Kane97}. Following Ref.\ \cite{Kane97}, we
find that a uniform $4^{o}$ twist of a $(10,10)$ tube
creates a gap of $0.6$ eV. The splay in the armchair lines
of Fig.\ \ref{fig4} may indicate a gentle bend of
the tube within the rope. As mentioned above, such bends
are predicted to have only a minor effect on the rope
resistivity.

Although a $4^{o}$ twist is large, it is consistent with
what is known about the mechanical properties of SWNTs.
Amazingly enough, SWNT mechanics can be modeled
rather well as if the tube were a continuous cylinder with
a wall thickness of $0.34$ nm, the interplane spacing
in graphite \cite{Lu97,Yakobson96}. A tube in a rope will
try as much as possible to align its hexagons with those of
neighboring tubes as in the AB stacking of graphite. In a
rope containing tubes of differing helicities, frustration
will induce tube twists. Countering this tendency is the
energy cost associated with an elastic twist distortion.
Little is known about the details of this energy balance
and possible energy barriers that define local energy
minima. Since ropes produced by laser ablation form within
a high temperature plasma that is rapidly quenched, there
could be substantial population of non-equilibrium
geometries. Large twists could occur even in ropes made
entirely of tubes with the identical wrapping vectors.

We now estimate the magnitude of twist that could be
induced by interactions between tubes of differing
helicities. Mentally unrolling the SWNT reveals that a
twist distortion of a tube is equivalent to a shear of the
underlying single plane of graphite (see Fig.\ \ref{fig3}b)
whose modulus we take from bulk graphite as \(M_{b} = 0.45\)
TPa. Similarly, a rotation of the tube away from its
equilibrium alignment with its neighbors in the rope likely
has a modulus similar to that of a shear distortion along
the graphite c-axis, \(M_{c} = 0.004\) TPa. For a $(10,10)$
tube, which is invariant under a rotation of $\pi/5$, the
maximum angular deviation from ideal alignment that can
occur is $\pi/10$. Since the tube radius is $0.7$ nm, the
equivalent shear for two graphite planes separated by
$0.34$ nm is about $e_{max} = 0.67$. The energy stored per
unit volume under a shear is the product of the relevant
modulus and the square of the shear. This implies that an
upper limit on the twist that can occur for the tube is \(
e_{max} \sqrt{M_{c}/M_{b}} = 0.067 \). Even larger twists
are possible if they are caused by included impurities in
the rope. Our observed twist of $4^{o}$ ($0.07$ rad) is
quite close to this estimated upper limit for twists
induced by tube-tube interactions alone.

It is very unlikely that STM artifacts are responsible for
the features we observe. We first consider artifacts due to
a finite tip diameter. An ideal zero-radius tip images the
projection of the tube on the substrate plane; a tip with
non-zero radius stretches the image perpendicular to the
axis of the circular tube. Distortions due to a symmetric
tip should be symmetric to the center axis of the tube and
not create an apparent twist. An asymmetric tip would lead
to an image where one side of the tube was stretched
compared to the other; again this would not be interpreted
as a twist.

Tube motion due to forces exerted by the STM tip may also
distort the image. Because the  tube surface is strongly
curved, we expect this force to be nearly perpendicular to
the tube axis. We have observed large tube shifts in our
experiments which typically appear as a sudden jump in the
image. Slower, more continuous tube motions that lead to
distorted images might also occur. To rule out this effect,
we took images of the region in Fig.\ \ref{fig4} at scan
angles as large as $15^{o}$ with respect to the tube axis.
They were essentially identical to those presented here.

In summary, we have obtained STM images that for the first
time allow the determination of the atomic structure of an
armchair SWNT with a twist distortion from its equilibrium
structure in isolation. Tube twist should be taken into
account when assigning wrapping vectors to SWNTs, and might
also occur in individual tubes that are strongly adsorbed on
substrates \cite{DekLieb}. Quenched twists are expected to
strongly affect the intrinsic electronic properties of
SWNTs, and they may contribute to the observed
insulator-like resistivity of SWNT ropes at low
temperature. Useful future experiments include quantifying
the twist frequency in SWNT ropes produced by different
techniques (laser ablation versus carbon arc), and
determining whether twist relaxation plays a role in
processing techniques (e.\ g.\, alkalai doping \cite{KMF})
known to increase the conductivity of SWNT material.

We thank the Smalley group of Rice University for providing
the SWNT material used in these experiments, J. Lefebvre for
assistance with the sample preparation, and C.L. Kane and
E.J. Mele for valuable discussions. This work was supported
by Penn's Laboratory for Research on the Structure of
Matter. W.C. thanks the Deutsche Forschungsgemeinschaft for
support. A.T.J. is supported by a David and Lucile Packard
Foundation Fellowship.

\begin{figure} 
\caption {Image of a rectangular SWNT rope on a Au(111)
substrate. Individual tubes are seen on the top of the
rope, along with single and multiple tube steps on the near
face.} 
\label{fig1} 
\end{figure} 
 
\begin{figure} 
\caption { a) High-resolution STM image of an area on top
of the rope. Hexagons of carbon atoms are clearly resolved.
The hexagon-hexagon distance is 0.25 nm, as expected for a
tube rolled from a single graphite sheet.   b) Line scan
perpendicular to the tube axis as shown by the white line
in the Fig.\ \ref{fig2}a.} 
\label{fig2} 
\end{figure}

\begin{figure} 
\caption {a ) Sketch of an undistorted armchair nanotube. \(
\mbox{\boldmath$ a_1$} \) and \( \mbox{\boldmath$ a_2$} \)
are the unit vectors of the graphite lattice. The armchair
and zig-zag directions are shown by solid and dotted lines,
respectively. b) Sketch of the atomic positions in a twisted
armchair tube. The armchair and zig-zag directions are
indicated as in Fig.\ 3a.} 
\label{fig3} 
\end{figure} 

\begin{figure} 
\caption {Image of the same area as in Fig.\ 2. The overall
tube curvature has been removed to increase the apparent
atomic corrugation. The dotted (solid) lines indicate the
zig-zag (armchair) directions, as in Fig.\ 3. Two grey
lines mark the edge of the tube.} 
\label{fig4} 
\end{figure} 
 
\end{document}